\RequirePackage{fix-cm}
\documentclass{svjour3}                     % onecolumn (standard format)
\smartqed  % flush right qed marks, e.g. at end of proof
\usepackage{graphicx}
\usepackage{amssymb}

\begin{document}

\title{Cryptanalysis and improvement of a quantum-communication-based online shopping mechanism %\thanks{Grants or other notes
}


\author{Wei Huang   \and
         Ying-Hui Yang  \and
         Heng-Yue Jia
}

\institute{W. Huang
        \at
State key Laboratory of Networking and Switching Technology, \\Beijing University of Posts and Telecommunications, Beijing 100876 China\\
              \and
              Y-H. Yang  \at
School of Mathematics and Information Science,\\Henan Polytechnic University, Jiaozuo 454000, China\\
\email{yangyinghui4149@163.com}
 \and
              H-Y Jia  \at
School of Information,\\Central University of Finance and Economics, Beijing 100081, China\\
%  \\
}
\date{Received: date / Accepted: date}

\maketitle

\begin{abstract}
Recently, Chou et al. [Electron Commer Res, DOI 10.1007/s10660-014-9143-6] presented a novel controlled quantum secure direct communication protocol which can be used for  online shopping. The authors claimed that their protocol was immune to the attacks from both external eavesdropper and internal betrayer. However, we find that this protocol is vulnerable to the attack from internal betrayer.  In this paper, we analyze the security of this protocol to show that the controller in this protocol is able to eavesdrop the secret information of the sender (i.e., the customer's shopping information), which indicates that it cannot be used for secure online shopping as the authors expected. Moreover, an improvement to resist the controller's attack is proposed.

\keywords{Quantum cryptography \and Quantum secure direct communication \and Cryptanalysis \and E-Commerce  \and Online shopping}
\end{abstract}

\section{Introduction}
\label{intro}

With the rapid development of the Internet and related technologies, E-commerce, which is one of the most significant scientific accomplishments brought by Internet, is playing an increasingly important role in modern life. As a key component of E-commerce, online shopping has become one of the most important shopping ways in people's everyday life. In 2013, the daily transaction volume of online shopping in the world reaches billions of dollars. Therefore, security and privacy have naturally become an essential requirement for online shopping.

So far, the security and privacy of E-commerce has been guaranteed by the classical cryptosystems whose security is based on the assumptions of computation complexity. Nevertheless, with the development of quantum algorithms and quantum computer \cite{1_Knill98,2_shor}, these classical cryptosystems are facing more and more challenges. To address the potential threat posed by quantum computation to classical cryptosystems, people begin to research new cryptographic technology, such quantum cryptography. Quantum cryptography, whose security is relied on the quantum mechanics principles rather than the assumptions of computation complexity, has become a hotspot of cryptography. Since the pioneering work of Bennett and Brassard in 1984\cite{3_BB84}, much attention has been focused on quantum cryptography, which includes quantum key distribution (QKD) \cite{3_BB84,4_DFG11,5_DFG12,6_Sun Y12,7_LinS14}, quantum secret sharing (QSS) \cite{8_HBB13,9_WTY16,10-GGP12,11_Schmid22,12_YFL34},  quantum secret direct communication (QSDC) \cite{13_LL33,14-LL34,15-Cai35,16-HW16,17_S.L37}, quantum watermark (QW) \cite{18_ZWW38,19_ZWW39,20_YYG40} and quantum secure multiparty computation (QSMC) \cite{21_GF41,22_ZKJ42,23_ZKJ43,24_WXJ44,25_YYG45,26_HT46,27_LYB47,28_ZWW48,29_HW49,30_HW50}, etc.

Quantum cryptography has also been utilized to assure the security and privacy in E-commence. In 2010, Wen presented an E-payment protocol by utilizing quantum group signature, in which a trusted third party is required \cite{31_WXJ51}. To enhance the robustness of the system,  Wen and Nie presented another E-payment protocol by employing quantum blind and group signature, where two trusted third parties are needed \cite{32_WXJ52}. Nevertheless, both the two protocols can only be applied to the cases where business transactions happen within the same bank. While in real life, many business transactions occur between different banks. In addition, an E-payment system, which  supports secure inter-bank transactions, should be desired from the view of practical application. In order to settle this problem and support unconditionally secure E-payment between two different banks, Wen et al. presented an inter-bank E-payment protocol based on quantum poxy blind signature in 2013 \cite{33_WXJ53}. Unfortunately, Cai and Wei found that this protocol is susceptible to denial-of-service attack. Moreover, they also show that the dishonest merchant can succeed to change the purchase information of the customer in this protocol \cite{34_WXJ54}.

It is known that design and cryptanalysis have always been important branches of cryptography. Both of them drive the development of this field. In fact, cryptanalysis is an important and interesting work in quantum cryptography \cite{35_gao55,36_WXJ56}. It estimates the security level of a protocol, finds potential loopholes, and tries to address security issues. As pointed out by Lo and Ko, \emph{breaking cryptographic systems was as important as building them} \cite{37-HKluo57}. To date, many kinds of attacks strategies have been presented, such as entanglement swapping attack \cite{38_LB38}, intercept-resend attack \cite{39-GaoPRL59}, Correlation-Extractability attack \cite{40-GF60,41-qsj61}, Trojan horse attack \cite{42-N.Gisin62} and participant attack \cite{43_GF63,44_GF64,45_GF65}.

Recently, Chou et al. presented a novel controlled QSDC protocol which can be used for online shopping \cite{46_GF65}. By utilizing this protocol, the online shopping mall could control the shopping process, hence  shopping information of the customer could be more secure.  For the sake of simplicity, we call it CLZ protocol hereafter. The authors made a simple security analysis of the CLZ protocol to show its immunity to the attacks from both external eavesdropper and internal betrayer. Unfortunately, we find that this protocol is susceptible to the attack from the internal betrayer. In this paper, we make an analysis to illustrate that this protocol cannot provide unconditional security for online shopping, since the controller of this protocol is able eavesdrop the secret information of the sender (i.e., the customer's shopping information). Moreover, we propose an improvement of the CLZ protocol to close this security loophole. In next section, we make a brief introduction of the CLZ protocol. In section 3, we make an analysis of the CLZ protocol to show its vulnerability to the internal attack from a dishonest controller. After that, we improve CLZ protocol to be secure against the presented attack.

\section{Brief review of the CLZ protocol and its application}

Herein we make a brief description of the CLZ protocol \cite{46_GF65}. Then we introduce how this protocol can be applied to online shopping.

\subsection{The CLZ protocol}

Different from the BB84 \cite{3_BB84} protocol in which only one photon is transmitted at a time, the photons in CLZ protocol are transmitted by utilizing the technique of block transmission which has been proposed firstly by Long et al. \cite{13_LL33}. In this protocol, Alice, Bob, and Charlie are supposed to be the information sender, receiver and controller, respectively. If Alice wants to transmit a secret message of $N$ bits directly to Bob under the control of Charlie, they could execute this protocol as the following steps.

\begin{enumerate}
\item[(1)] \emph{Controller prepares for the quantum information carriers}. Charlie prepares a sequence of  $N$+$\delta$ single qubits, each of which is randomly in one of following four states
\{$|0\rangle$, $|1\rangle$, $|+\rangle$, $|-\rangle$\}, where
\begin{eqnarray}
    |+\rangle=\frac{1}{\sqrt{2}}(|0\rangle+|1\rangle),\quad\quad\quad\;|-\rangle=\frac{1}{\sqrt{2}}(|0\rangle-|1\rangle),
\end{eqnarray}
and the sequence is denoted as $S_0$. After that,  he sends  $S_0$ to Alice.
\item[(2)]\emph{Eavesdropping check for the first transmission}. After the reception of $S_0$, Alice begins to check eavesdropping with Charlie as follows. Alice randomly selects $\delta$ single qubits from $S_0$ and randomly chooses a measuring basis $Z$-basis or $X$-basis, measuring the selected qubits for checking eavesdropping. Afterwards,  Alice informs  Charlie of the information including which bases she uses, the positions of selected qubits and the corresponding measurement outcomes. With the information from Alice, Charlie could determine whether there exists eavesdropping in the first transmission. If there exists eavesdropping,  they abort the protocol; otherwise, Alice throws away the qubits used for checking eavesdropping and continue to the next step.
\item[(3)]\emph{Sender encodes the secret information}. Alice encodes her secret message $M$ on the remaining $N$ qubits as follows. If the $i$-th bit of $M$ is 0/1, Alice performs operation $I$/$i$$\sigma_y$ on the $i$-th one of the remaining qubits, where
\begin{eqnarray}
   I=|0\rangle\langle0|+|1\rangle\langle1|,\quad\quad i\sigma_y=|0\rangle\langle1|-|1\rangle\langle0|.
\end{eqnarray}
    After that, she generates $\delta$ decoy qubits which are randomly in one of the four states in \{$|0\rangle$, $|1\rangle$, $|+\rangle$, $|-\rangle$\} and inserts them randomly into the sequence $S_0$. Then Alice send the new sequence (denoted as $S_1$) to Bob.
\item[(4)] \emph{Eavesdropping checking for the second transmission}. After Bob receives $S_1$ from Alice, Alice informs Bob of the positions of the decoy qubits, Bob measures each of the decoy qubits randomly in $Z$-basis or $X$-basis. After that, Bob tells Alice which bases he uses and the corresponding measurement outcomes. According to the information announced by Bob, Alice could determine whether there exists eavesdropping during the transmission of $S_1$ . If there exists eavesdropping,  they stop the protocol; otherwise, Bob could get the secret message with the help of Charlie as follows.
\item[(5)]\emph{Receiver deduces the secret message with the controller's help}. Bob discards the $\delta$ qubits used for checking eavesdropping and now only remains $N$ single qubits. Without Charlie's permission, Bob is unable to obtain Alice¡¯s secret message. Only after Charlie publishes the initial states of the $N$ single qubits, can Bob recover Alice's secret message $M$ by comparing with the initial states. Concretely, if the initial state of a qubit is $|0\rangle$ or $|1\rangle$ ($|+\rangle$ or $|-\rangle$), Bob measure it in $Z$-basis ($X$-basis). And the corresponding bit of $M$ is 0 (1) provided the measurement outcome is the same as (different with) the initial state.
\end{enumerate}

This proposed protocol could also be extended to a multiparty controllers version. Take the two controllers version as a example. Suppose the two controllers, Charlie and Dave, both have the ability to control Alice and Bob's communication. Some modifications are required in step (1) and step (5). In step (1),  after Charlie prepares the qubit sequence, he sends it to Dave instead of Alice. Upon Dave receives these photons, he make an eavesdropping check as Alice does in step (2). To change the states of qubits, Dave performs randomly operation $I$ or $i$$\sigma_y$  on each of the remaining qubits, then he sends these qubits to Alice like original step (1) does. In the new step 5, Bob need ask Charlie to reveal the initial states he prepared and Dave to publish the operations he had performed. After Bob gets all the information, he is able to deduce $M$.

\subsection{The application of the CLZ protocol on online shopping}

Herein we introduce how the CLZ protocol can complete the online shopping process \cite{46_GF65}. In this example,  eBay is considered as the online shopping
mall, and the detail steps can be described as follows.

\begin{enumerate}
\item[(a)]Both seller and costumer register as eBay members.
\item[(b)]eBay authenticates the identities of seller and costumer.
\item[(c)]Once the customer decides to buy items from the seller, he/she asks eBay to transmit a sequence of single qubits randomly in \{$|0\rangle$, $|1\rangle$, $|+\rangle$, $|-\rangle$\} via quantum channel, but without knowing its initial states.
\item[(d)]After checking the security of the transmitted qubits, customer encodes his/her shopping information on those photons by performing the corresponding unitary operations as described in steps (2) and (3). The shopping information includes customer ID, item number, etc. Once the encoding is finished, those encoded photons will be sent to the seller.
\item[(e)]Again, the security checking on these transmitted qubits is needed. If the transmission is secure, seller will ask eBay for the initial states of these photons. With the encoded qubits and their initial states, seller could deduce the shopping information of the customer.
\end{enumerate}

\section{Security analysis and improvement of the CLZ protocol}
In this section, we analyze the security of the CLZ protocol to show that it is susceptible to attack from a dishonest controller. Specifically, we first explain how a dishonest controller could eavesdrop the secret information of the sender in the CLZ protocol. Then we illustrate that, by utilizing this strategy, a dishonest online shopping mall can eavesdrop the customer's shopping information in the application of online shopping. Finally, we introduce our improvement of the CLZ protocol to close the corresponding security loophole.

\subsection{Security analysis}
In Ref. \cite{46_GF65}, the authors make a simple analysis to show that the CLZ protocol is secure against the attacks from both external eavesdropper and internal betrayer. In each transmission of the quantum information carriers in the CLZ protocol, there are $\delta$ decoy qubits, which are randomly in the four states $\{|0\rangle, |1\rangle, |+\rangle, |-\rangle\}$ and randomly inserted in the qubit sequence. After receives the qubit sequence, the receiver measures each of the decoy qubits randomly in $Z$-basis or $X$-basis.  Whatever kind of attack an external eavesdropper utilizes, his/her eavesdropping action will  introduces disturbance into the eavesdropping check with a certain probability.  The reason is that the process for eavesdropping check done in this protocol, in essence, is the same as that in the BB84 QKD protocol \cite{3_BB84}, which has been proved unconditional secure. Hence, the CLZ is indeed secure against the external attacks.

However, an internal betrayer of a multiparty quantum cryptographic protocol may has more power to attack the protocol than an external eavesdropper. First, he/she can know partial information legally. Second, he can tell a lie in the process of eavesdropping check to avoid introducing errors. Therefore, the attacks from dishonest participants are generally more powerful and should be paid more attention to \cite{23_ZKJ43,24_WXJ44,25_YYG45}. The authors of the CLZ protocols said that this protocol could resist the attacks from both the information receiver and the controller. That is, the receiver Bob could not get the secret message $M$ without the help of the controller Charlie. Also Charlie is unable to get $M$ without leaving a trace in the eavesdropping check.

We admit that, no matter what kind of attack Charlie employs in the CLZ protocol, once he could get partial information of $M$, her action will introduce errors into the eavesdropping check and hence make the protocol aborted. In fact, this condition is sufficient to ensure the security of a QKD protocol, but it is not enough to guarantee the security of the CLZ protocol since it is a QSDC protocol. Different from QKD, the purpose of which is to establish a private random key, the purpose of QSDC is to directly transmit a secret message \cite{13_LL33,14-LL34,15-Cai35,16-HW16,17_S.L37}. In a QKD protocol, if the eavesdropper's action is detected in the eavesdropping check, the transmitted qubits can be abandoned as they do not carry any information about the secret message. On the contrary, in a QSDC protocol, the secret message is directly encoded on the transmitted qubits. Hence, QSDC has higher security requirements than QKD. On the one hand, the eavesdropping check in a secure QSDC protocol should detect the eavesdropper's attack. On the other hand, a secure QSDC protocol should leak none useful information of the secret message to the eavesdropper. In other words,  once the eavesdropper has already gotten any useful information of the secret message, the QSDC protocol is insecure even if the eavesdropper's attack has been detected in the eavesdropping check.

Unfortunately, the CLZ protocol could not simultaneously satisfy both the two security requirements. In this protocol, if Charlie wants to eavesdrop the secret message $M$, he could execute the following strategy. In step (3), when Alice sends out the sequence $S_1$, Charlie intercepts the travelling sequence $S_1$ and stores it. Instead, he prepares another sequence $S'_1$ of $N$+$\delta$ qubits to replace $S_1$ and sends it Bob. After Bob receives the sequence $S'_1$, Alice will announce the positions of the $\delta$ decoy qubits. Once Charlie gets this information, he discards the corresponding decoy qubits in $S_1$. Then he could easily deduce $M$ with the remaining $N$ qubits and their initial states. It should be pointed out this attacking  strategy will inevitably introduce errors into the eavesdropping check. Thus, the protocol will be aborted by Alice and Bob according to step (4). Even so, Alice and Bob could only determine that there exists eavesdropping in the transmission. That is to say, they will not suspect Charlie as the eavesdropper. So far, we have shown that the controller Charlie in the CLZ protocol could easily get the whole secret message $M$ beyond suspicion. Accordingly, the application of the CLZ protocol given above, i.e., online-shopping, also have the same security loophole. Specifically, the online shopping mall (i.e., eBay) could utilize this strategy to get the customer's shopping information in the application of online shopping.

\subsection{Improvement of the CLZ protocol}

Herein we give an improvement of the CLZ protocol in order to close the security loophole introduced above. To close the corresponding loophole, we only need to respectively substitute steps (3)-(5) of the  CLZ protocol with the steps (3')-(5') given below.

\begin{enumerate}
\item[(3')]\emph{Sender encodes the secret information}. Before encoding her secret information, Alice first generates an $N$-bit random binary string $K$. Then she encodes $K$$\oplus$$M$ on the remaining $N$ qubits as follows (Here $\oplus$ represents XOR operation). If the $i$-th bit of $K$$\oplus$$M$ is 0/1, Alice performs operation $I$/$i$$\sigma_y$ on the $i$-th one of the remaining qubits. After that, she generates $\delta$ decoy qubits which are randomly in one of the four states in \{$|0\rangle$, $|1\rangle$, $|+\rangle$, $|-\rangle$\} and inserts them randomly into the sequence $S_0$. Then Alice sends the new sequence (denoted as $S_1$) to Bob.
\item[(4')] \emph{Eavesdropping checking for the second transmission}. After Bob receives $S_1$ from Alice, Alice informs Bob of the positions of the decoy qubits, Bob measures each of the decoy qubits randomly in $Z$-basis or $X$-basis. After that, Bob tells Alice which bases he uses and the corresponding measurement outcomes. According to the information announced by Bob, Alice could determine whether there exists eavesdropping during the transmission of $S_1$ . If there exists eavesdropping,  they stop the protocol; otherwise, Alice publishes the binary string $K$. Then Bob could get the secret message with the help of Charlie and $K$ in the next step.
\item[(5')]\emph{Receiver deduces the secret message with the controller's help}. Bob discards the $\delta$ qubits used for checking eavesdropping and now only remains $N$ single qubits. Without Charlie's permission, Bob is unable to obtain Alice¡¯s secret message. Only after Charlie publishes the initial states of the $N$ single qubits, can Bob recover the binary string $K$$\oplus$$M$ by comparing with the initial states. Concretely, if the initial state of a qubit is $|0\rangle$ or $|1\rangle$ ($|+\rangle$ or $|-\rangle$), Bob measure it in $Z$-basis ($X$-basis). And the corresponding bit of $K$$\oplus$$M$ is 0 (1) provided the measurement outcome is the same as (different with) the initial state. Once obtaining $K$$\oplus$$M$, Bob could deduce $M$ with the string $K$ announced by Alice.
\end{enumerate}

Now we show that this improvement can be used to close the above security loophole of the CLZ protocol. As analyzed in section 3.1, the  process for eavesdropping check done in the CLZ protocol, in essence, is the same as that in the BB84 QKD protocol \cite{3_BB84}. Therefore, no matter what kind of attack the eavesdropper (both the external eavesdropper and internal betrayer) utilizes, once he/she has obtained any useful information about $K$$\oplus$$M$, he/his attacking action will unavoidably leave a trace in (i.e., introduce errors into ) the check. Then the protocol will be stopped  before Alice publish the binary string $K$. Since $K$ is a random binary string known only by Alice, even if the eavesdropper get any useful information about $K$$\oplus$$M$, he/she knows nothing about $M$ since $K$ will not be published anymore. Till now, we have shown that the improved CLZ protocol is secure against all the present attack, since whatever kind of attack the eavesdropper uses, he/she could get none useful information about $M$ but be noticed in the eavesdropping check. Accordingly, by utilizing the improved version, the application of the improved CLZ protocol in online shopping could also be immune to all the present attack.

\section{Conclusion}
In this paper, we make a cryptanalysis of the CLZ protocol, which can be used for online shopping, to show that it has a security loophole. Concretely, we point out that the controller of the CLZ protocol is able to obtain the whole secret message of the sender. Then, we improve this protocol to be secure against all the present attacks.

\begin{acknowledgements}
This work is supported by NSFC (Grant Nos. 61300181, 61272057, 61202434, 61170270, 61100203, 61121061,61309029), Beijing Natural Science Foundation (Grant No. 4122054), Beijing Higher Education Young Elite Teacher Project (Grant Nos. YETP0475, YETP0477), BUPT Excellent Ph.D. Students Foundation (Grant No. CX201441).
\end{acknowledgements}



\end{document}